**A Framework for Characterizing Learning Contributions Across the Initial-Achievement Spectrum**

Tianlong Zu

*Department of Physics and Astronomy, Northwestern University, 2145 Sheridan Road, Evanston, IL, USA 60208*

**Abstract:** Conceptual assessments are widely used in physics education research to evaluate changes in student understanding, yet class-average measures can obscure how those changes are distributed across students with different levels of initial achievement. We introduce a Learning Contribution Framework that characterizes this distribution through the Learning Contribution Curve (LCC) and Learning Contribution Profile (LCP). For a general contribution measure $G$, the LCC represents cumulative contribution across students ranked by initial achievement, whereas the LCP describes the local mean contribution relative to the population mean of the individual $G$ values. We apply the framework to the individual Hake normalized gain and develop a statistical model linking LCC and LCP behavior to the joint structure of pretest and posttest scores. Under the central assumption that the conditional mean of the subsequent score is linear in the initial score, we identify a score-structure parameter $\beta$ and a critical score-structure parameter $\beta_c$. Whether $\beta$ is greater than, less than, or equal to $\beta_c$ determines whether the expected LCP increases, decreases, or remains constant across initial achievement. Given the pretest distribution, the model further yields analytical predictions for the complete LCP and LCC that closely reproduce simulation results. Application to classroom concept-assessment data illustrates how the framework reveals local and cumulative patterns of learning contribution that are not evident from an overall class average revealed by the traditional Hake's normalized gain.

## 1. Introduction

An important goal of physics education research (PER) is to develop and evaluate instructional approaches that improve students' conceptual understanding of physics. To support this effort, the PER community has developed and widely adopted research-validated conceptual assessments, including the Force Concept Inventory (FCI), The Force and Motion Conceptual Evaluation (FMCE), the Brief Electricity and Magnetism Assessment (BEMA), the Determining and Interpreting Resistive Electric Circuits Concepts Test (DIRECT), and numerous other instruments [1-5]. When administered before and after instruction, these assessments provide a means of quantifying changes in conceptual understanding and comparing learning outcomes across instructional settings.

One of the earliest and most influential measures for characterizing such changes is the normalized gain, introduced by Hake [6]. The class-average normalized gain represents the observed improvement from pretest to posttest relative to the maximum possible improvement given the class's initial performance. Hake's analysis of more than 6000 students in introductory mechanics courses found substantially larger normalized gains in courses employing interactive-engagement methods than in courses using traditional instruction [6], establishing normalized gain as a widely used measure of conceptual learning in PER.

Subsequent work has examined both the properties and limitations of normalized gain $g$-factor. Bao showed that normalized gain calculated from class-average scores is generally not equivalent to the arithmetic mean of students' individual normalized gains and that the difference between these quantities contains information about the relationship between initial performance and individual learning gains [7]. Hake subsequently examined individual normalized gains and their relationships with students' initial conceptual understanding and other measures of preparation [8]. Nissen and colleagues compared normalized gain with Cohen's $d$ and demonstrated that the choice of learning metric can affect conclusions concerning learning and educational equity [9]. More recently, Navarrete and collaborators interpreted normalized gain within a learning-rate model and examined the effects of measurement error on individual- and group-level estimates of learning [10]. They showed that measurement error can induce an apparent relationship between observed pretest scores and individual normalized gains even when the underlying learning rate is independent of initial knowledge. Their analysis also highlighted the importance of aggregation when estimating learning at the group level.

These studies establish normalized gain as a useful measure of overall learning while also highlighting an important limitation of any single class-level statistic: it cannot describe how learning is distributed across students with different levels of initial achievement. Several studies have therefore examined learning outcomes as a function of students' initial understanding. Pawl, for example, investigated individual FCI gains as a function of pretest score and showed that a class-average normalized gain can provide an incomplete description of individual student learning [11]. Such analyses illustrate that students entering the same instructional environment with different levels of initial preparation may exhibit substantially different learning outcomes. A related development is the conceptual growth curve introduced by Christman, Miller, and Stewart [12]. Using a large set of matched FMCE pretest and posttest data, they modeled posttest performance as a function of pretest performance. This approach characterizes expected post-instruction performance across different levels of initial preparation and facilitates comparisons between instructional populations that may differ systematically in incoming preparation. Other research further suggests that instructional approaches may not affect all students uniformly. Traditional introductory physics courses may disproportionately benefit students who enter with stronger preparation, whereas appropriately designed active-learning environments have, in some contexts, reduced failure rates and achievement disparities among students with weaker preparation or greater academic risk [13]. These effects are not universal, however, and may depend substantially on the design and implementation of the instructional environment [14]. Together, these findings suggest that instructional effectiveness cannot be characterized fully by class-average outcomes alone and motivate closer examination of how observed learning gains are distributed across the initial-achievement spectrum.

Taken together, this body of work reflects a progression from class-average learning gains to individual learning gains and their relationship with initial performance, to explicit modeling of the relationship between pretest and posttest performance. These developments increasingly retain information that is lost when an entire pretest-posttest dataset is reduced to a single summary statistic. Nevertheless, a complementary distributional question remains:

"*How is the total observed learning gain concentrated across students with different levels of initial achievement?*"

This question focuses on information that is largely lost when learning outcomes are summarized by a single class-average measure. Class-average normalized gain provides an overall measure of improvement, but it does not show how that improvement is distributed across students with different levels of initial achievement. Consider two classes with the same overall normalized gain. In one class, students across the initial-achievement spectrum may contribute relatively evenly to the total gain. In another, a disproportionate share of the gain may come from students entering with relatively low conceptual understanding, while students with higher initial achievement exhibit smaller or even negative gains. Although the two classes are indistinguishable in terms of their average normalized gain, they exhibit qualitatively different distributions of individual learning change across initial achievement. Characterizing this distribution therefore provides information about class performance that is not captured by the class average alone.

To characterize this distributional structure, we introduce a learning contribution framework consisting of two complementary representations: the Learning Contribution Curve (LCC) and the Learning Contribution Profile (LCP). Together, they describe how learning gains are distributed across students with different levels of initial achievement, both cumulatively and locally.

The LCC provides the cumulative representation. Students are ordered from lowest to highest according to pretest performance, and the cumulative fraction of total individual learning gain is plotted against the cumulative fraction of students. The LCC therefore shows how learning accumulates across the initial-achievement distribution rather than reducing that distribution to a single class-average value. The LCP provides the corresponding local representation. Defined by the slope of the LCC, the LCP expresses the average learning gain within a local region of the initial-achievement distribution relative to the overall average individual gain. An LCP value greater than unity indicates above-average learning contribution, a value between zero and unity indicates positive but below-average contribution, and a negative value indicates negative average learning gain. The LCC and LCP therefore form complementary components of the learning contribution framework, with the LCC describing the cumulative distribution of learning contribution and the LCP describing its local variation across initial achievement. The LCC is mathematically related to rank-dependent concentration curves, including the Lorenz curve [15, 16, 17]. Unlike a conventional Lorenz curve, however, students are ranked by initial achievement while a different and potentially signed quantity, learning contribution, is accumulated. Consequently, the LCC need not be monotonic, convex, or restricted to one side of the equal-contribution line.

The purpose of the present work is not to replace class-average normalized gain, but to provide a complementary framework for characterizing how learning gains are distributed across the initial-achievement spectrum. We first delineate the general learning-contribution framework and define the LCC and LCP for a generic measure of learning contribution. We then specialize the framework to individual normalized gain and examine the resulting mathematical properties and theoretically distinct contribution patterns. Next, we apply the framework to pretest–posttest data from a research-based physics concept inventory to illustrate how it can reveal features of student learning that are not apparent from class-average measures alone. Finally, we discuss the limitations of the present formulation and possible directions for extending the framework in future work.

## 2. The Learning Contribution Framework

### 2.1 The general mathematical foundation

Consider a class of $N$ students for whom an initial-achievement measure $P$ and a corresponding measure of learning change $G$ are available. Conventional class-level statistics summarize the overall magnitude of change but do not indicate how that change is distributed among students entering with different levels of initial achievement. To characterize this distributional structure, we introduce the LCC.

We begin with a general formulation. Let $G_i$ denote a quantitative learning change of student $i$. The specific definition of $G_i$ is not intrinsic to the LCC framework; rather, it is chosen according to the assessment and research question. Students are ordered from lowest to highest according to their initial achievement, $P_1 \le P_2 \le \cdots \le P_N$. For the first $k$ students in this ordering, define

$$x_k = \frac{k}{N}, \qquad k = 0, 1, \dots, N \tag{1}$$

and

$$y_k = \frac{\sum_{i=1}^{k} G_i}{\sum_{i=1}^{N} G_i} = \frac{\sum_{i=1}^{k} G_i}{N\bar{G}} \tag{2}$$

Throughout the present work, we consider the case in which the mean individual contribution is strictly positive,

$$\bar{G} = \frac{1}{N}\sum_{i=1}^{N} G_i > 0 \tag{3}$$

With $x_0 = y_0 = 0$, the resulting curve begins at $(0, 0)$ and terminates at $(1, 1)$. A point on the LCC has a direct cumulative interpretation: the lowest $100x\%$ of students, ranked according to initial achievement, account for $100y\%$ of the total net contribution measured by $G$.

Note the slope of the LCC possesses a meaningful local interpretation. Note that for consecutive students,

$$\Delta x_k = x_k - x_{k-1} = \frac{1}{N} \tag{4}$$

and

$$\Delta y_k = y_k - y_{k-1} = \frac{G_k}{N\bar{G}} \tag{5}$$

Therefore, the slope about position $x_k$ can be calculated as:

$$s_k = \frac{\Delta y_k}{\Delta x_k} = \frac{NG_k}{N\bar{G}} = \frac{G_k}{\bar{G}} \tag{6}$$

Thus, the slope of the LCC has a direct interpretation: it measures the contribution associated with a student $k$, relative to the mean individual contribution of the class. A slope of unity corresponds to a contribution equal to average contribution. A slope greater than unity indicates an above-average contribution, whereas a slope between zero and unity indicates a positive but below-average contribution. A negative slope indicates a negative contribution. We will refer the slope of LCC as LCP.

In a continuous population representation where $N \to \infty$ and $\Delta x = 1/N \to dx$. The population LCC may then be written using calculus:

$$y(x) = \frac{1}{\bar{G}} \int_0^x G(u)du \tag{7}$$

Note that $0 \le x \le 1$ and $\bar{G} = \int_0^1 G(x)dx$ where $G(x)$ can be treated as a continuous function of $x$. This representation makes explicit that the LCC accumulates the local individual contribution as one moves through the initial-achievement distribution.

The corresponding local representation is obtained by differentiating the LCC. We define the LCP as

$$s(x) = \frac{d}{dx} y(x) \tag{8}$$

Using the expression above,

$$s(x) = \frac{G(x)}{\bar{G}} \tag{9}$$

The LCP therefore expresses the mean contribution associated with students near rank $x$ relative to the population mean contribution. The normalization of the LCC requires $y(0) = 0$ and $y(1) = 1$. Therefore,

$$\int_0^1 s(x)dx = 1 \tag{10}$$

The horizontal line $s(x) = 1$ thus provides a natural reference for uniform contribution across the initial-achievement distribution. Equivalently, if the mean contribution is independent of initial-achievement rank, then throughout the population and the corresponding LCC is $y(x) = x$. We refer to as the equal-contribution line.

**2.2 Treatment of tied ranking scores**

Because conceptual assessments have discrete score scales, multiple students may receive the same score on the variable used to rank the class. In such cases, the relative ordering of students within a tied-score group is undefined. If students sharing the same pretest score were ordered arbitrarily, their individual learning-contribution values $G_i$ could appear in different sequences, producing different intermediate paths in the LCC and different apparent local patterns in the LCP. To eliminate this arbitrary dependence, we adopt a tie-invariant construction in which students with the same pretest score are treated collectively.

Let the class have $K$ tied-score groups, ordered by increasing pretest score and indexed by $j = 1, 2, \dots, K$. Note $K$ cannot be greater than the number of items on a concept assessment. Let $n_j$ denote the number of students in tied-score group $j$. These students therefore occupy ranks: $k+1, k+2, \dots, k+n_j$. Note there are a total of $k = \sum_{i=0}^{j-1} n_i$ students in the groups 1 through group $j-1$, we characterize the tied group $j$ by its mean learning contribution

$$\bar{G}_{n_j} = \frac{1}{n} \sum_{i=k+1}^{k+n_j} G_i \tag{11}$$

For construction of the tie-invariant LCC, each student $i$ within this tied block is assigned the same arithmetic average of the individual measure $G_i^* = \bar{G}_{n_j}$. The tied block therefore begins at the existing LCC point $(x_k, y_k)$, and ends at $(x_{k+n_j}, y_{k+n_j})$. Because $x_k = k/N$, and $x_{k+n_j} = (k+n_j)/N$, the horizontal width of the tied block is

$$\Delta x_{n_j} = x_{k+n_j} - x_k = \frac{n_j}{N} \tag{12}$$

Thus, the horizontal length of the corresponding LCC and LCP segments is directly proportional to the number of students in the tied-score group. And the cumulative contribution at the end of the tied block is

$$y_{k+n_j} = y_k + \frac{n_j}{N} \frac{\bar{G}_{n_j}}{\bar{G}} \tag{13}$$

Equivalently, intermediate points within the tied block lie on the straight line connecting $(x_k, y_k)$ and $\left(x_{k+n_j}, y_{k+n_j}\right)$, ensuring that the resulting LCC and LCP are independent of any arbitrary ordering of students with identical ranking scores. The slope across this interval is:

$$s_{n_j} = \frac{y_{k+n_j} - y_k}{x_{k+n_j} - x_k} = \frac{\bar{G}_{n_j}}{\bar{G}} \tag{14}$$

The corresponding LCP is therefore constant over the interval $x_k < x_i \le x_{k+n_j}$. Thus, in the tie-invariant LCP, the width of each horizontal segment represents the fraction of students sharing the same ranking score, its height represents that group's mean learning contribution relative to the average individual contribution. Consequently, the tie-invariant construction of LCP offers a local tied-score group level interpretation:

1) $s_{n_j} > 1$ indicates a region in which the local tied-score group mean contribution exceeds the mean individual contribution.
2) $0 < s_{n_j} < 1$ indicates a positive but below-average local tied-score group contribution.
3) $s_{n_j} = 1$ indicates a local tied-score group learning contribution equal to the mean individual contribution.
4) $s_{n_j} < 0$ indicates a region with a negative local contribution.

**2.3 Application to individual normalized gain**

The general construction above does not require a particular definition of $G_i$. In the present study, we specialize the framework to individual Hake normalized gain $g_i$. Let $P_i$ and $Q_i$ denote the pretest and posttest scores of student $i$, respectively, and let $M$ denote the maximum possible score. For $P_i < M$, we choose $G_i$ to be the individual $g$-factor:

$$G_i = g_i = \frac{Q_i - P_i}{M - P_i} \tag{15}$$

The quantity $g_i$ represents the observed score improvement as a fraction of the maximum improvement available from the student's initial score. Individual values may be positive, zero, or negative. With this choice, the LCC becomes

$$x_k = \frac{k}{N}, \quad y_k = \frac{\sum_{i=1}^{k} g_i}{\sum_{i=1}^{N} g_i} = \frac{\sum_{i=1}^{k} g_i}{N\bar{g}} \tag{16}$$

Where $\bar{g}$ is the overall mean individual normalized gain. And the LCP becomes:

$$s_k = \frac{g_k}{\bar{g}} \tag{17}$$

Note the population LCC can be expressed as

$$y(x) = \frac{1}{\bar{g}} \int_0^x g(u)du \tag{18}$$

and the population LCP can be expressed as

$$s(x) = \frac{g(x)}{\bar{g}} \tag{19}$$

Where $\bar{g} = \int_0^1 g(x)dx$. In the continuous limit (large class size), the slope therefore represents the relative learning effectiveness of students near a given pretest percentile.

Using the tied-score group formalism introduced in Sec. 2.2, we can develop the tie-invariant construction of the LCC and LCP after the $g$-factor is adopted. The average individual normalized gain of the tied-score group $j$ is

$$\bar{g}_{n_j} = \frac{1}{n_j} \sum_{i=k+1}^{k+n_j} g_i \tag{20}$$

The LCC within the tied-score interval is defined by assigning this group-average gain uniformly across the $n_j$ student positions. Accordingly, the LCC is linear across the tied-score interval:

$$y_{k+m} = y_k + \frac{m}{N} \frac{\bar{g}_{n_j}}{\bar{g}}, \qquad m = 0, 1, \dots, n_j \tag{21}$$

And the corresponding LCP is constant, with value:

$$s_{n_j} = \frac{\bar{g}_{n_j}}{\bar{g}} \tag{22}$$

This construction ensures that both the LCC and LCP are independent of any arbitrary ordering of students within the tied group.

We could also demonstrate that, for students sharing the same pretest score $P_i$, the mean individual normalized gain $\bar{g}_{n_j}$ is mathematically identical to the normalized gain calculated from the tied-score group mean pretest and posttest scores.

$$\bar{g}_{n_j} = \frac{1}{n_j} \sum_{i=k+1}^{k+n_j} \left( \frac{Q_i - P_i}{M - P_i} \right) = \frac{\left( \sum_{i=k+1}^{k+n_j} Q_i / n_j \right) - P_i}{M - P_i} \tag{23}$$

Since the mean tied-score group posttest and pretest scores are

$$\bar{Q}_{group\ j} = \frac{1}{n_j} \sum_{i=k+1}^{k+n_j} Q_i, \qquad P_i = \bar{P}_{group\ j} \tag{24}$$

It follows that,

$$\bar{g}_{n_j} = \frac{\bar{Q}_{group\ j} - \bar{P}_{group\ j}}{M - \bar{P}_{group\ j}} \tag{25}$$

Thus, the tie-invariant local gain used in the LCP can equivalently be interpreted as the Hake normalized gain calculated from the mean pretest and posttest scores of the tied-score group. This equivalence provides a direct local tied-score group-level interpretation of the LCP, $s_{n_j} = \bar{g}_{n_j}/\bar{g}$, as the normalized gain of the local tied-score group relative to the overall mean individual normalized gain of the class.

Note that individual normalized gain, $g_i$, represents an observed change in assessment performance and may include both learning and student-level measurement variation. The LCP therefore does not interpret each $g_i$ as a precise measure of latent learning. Instead, it characterizes the mean observed gain among students with similar or identical initial achievement. Averaging in this manner reduces the influence of idiosyncratic fluctuations, although systematic biases associated with initial achievement may remain. Accordingly, the LCP should be interpreted as a profile of local average observed gain: $s(x) > 1$ indicates regions with above-average normalized gain, whereas $s(x) < 1$ indicates regions with below-average gain.

Although individual normalized gain is used in the present analysis, it represents only one possible choice of the learning-contribution measure $G$. The same framework can, in principle, be applied to raw score gain, or any other individual-level change estimate. appropriate to the research question. The general construction and interpretation of the LCC and LCP apply across such choices of $G$, whereas some of the specific mathematical properties developed in this work arise from the adoption of the normalized gain.

It is tempting to interpret the equal-contribution line, $y = x$, as an equity reference. Such an interpretation requires caution, however, because the framework itself does not impose a particular definition of educational equity. The line $y = x$, equivalently $s(x) = 1$, represents only equal learning contribution with respect to the measure $G$ used in the analysis. When $G$ is individual normalized gain, this condition corresponds to equal proportional realization of the available score improvement across the initial-achievement distribution. Whether such a pattern should be considered equitable depends on the equity criterion being adopted. The LCC and LCP can therefore inform equity-related analyses by making distributional patterns visible, but those patterns should not be interpreted as evidence of equity or inequity without an explicitly defined criterion.

## 3. Illustrative Models

To illustrate the range of distributional patterns that can arise within this framework, we next consider a set of idealized population-level LCPs. Rather than beginning with a particular empirical dataset, we specify representative functional forms for $s(x)$ that describe different ways in which learning contributions may vary across the initial-achievement distribution. The corresponding LCCs are then obtained from Eq. (18) subject to the normalization condition expressed via Eq. (10). Five hypothetical models spanning several qualitatively distinct behaviors are discussed below and the model LCCs and LCPs are presented in Fig. 1.

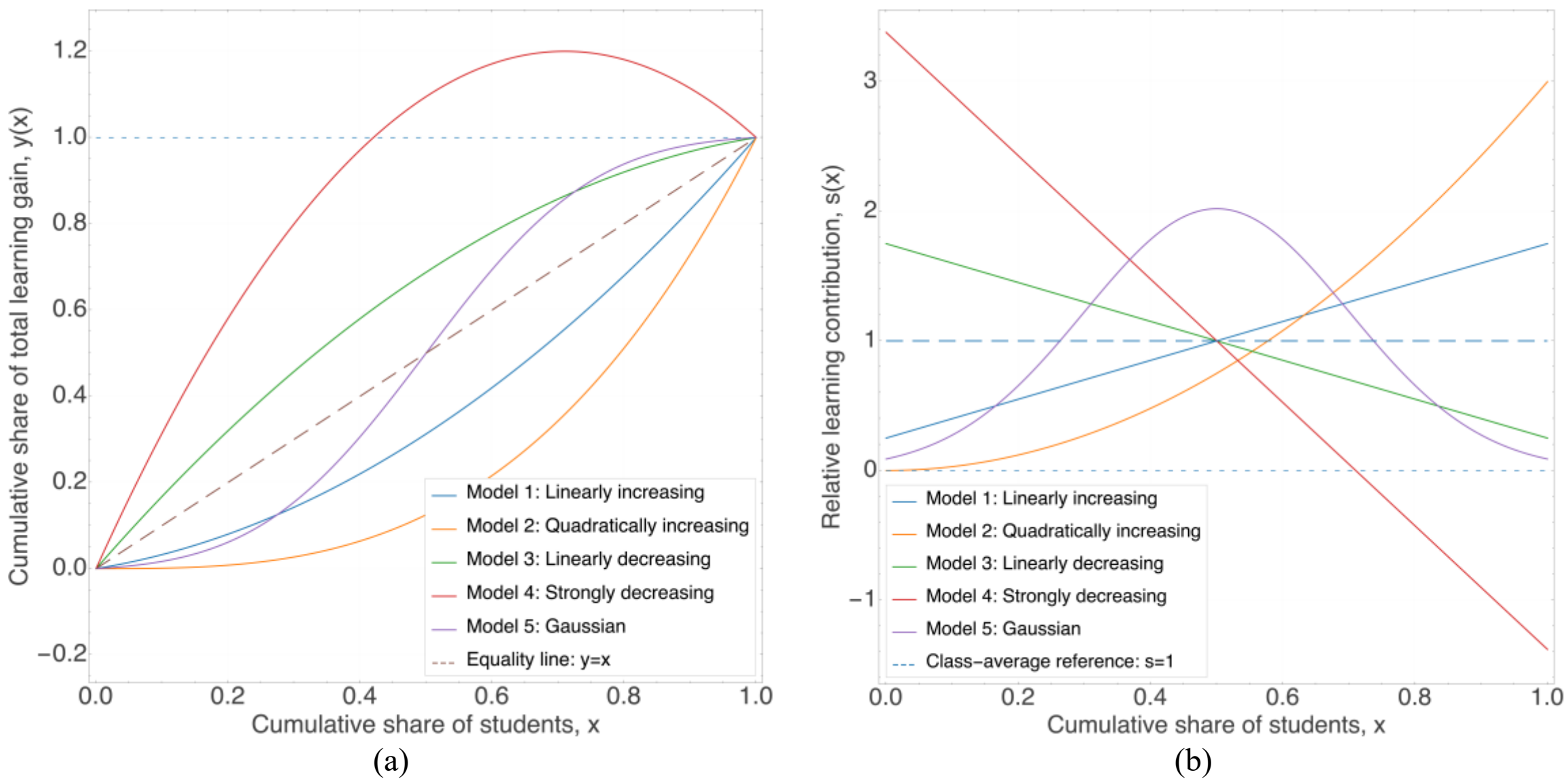

**FIG. 1.** Five illustrative learning-contribution patterns. (a) LCCs and (b) corresponding LCPs.

### a) Linearly increasing contribution.

The slope is assumed to increase linearly with the ranking score,

$$s(x) = 0.25 + 1.5x \tag{26}$$

Integration gives

$$y(x) = 0.25x + 0.75x^2 \tag{27}$$

Because $s(x) < 1$ for $x < 0.5$ and $s(x) > 1$ for $x > 0.5$, students in the lower half of the initial-achievement distribution exhibit below-average contribution, whereas those in the upper half exhibit above-average contribution. The resulting LCC is convex and lies below the equal-contribution line.

### b) Quadratically increasing contribution.

The second model assumes that relative learning contribution increases quadratically and begins at zero initially,

$$s(x) = 3x^2 \tag{28}$$

The corresponding LCC is

$$y(x) = x^3 \tag{29}$$

In this model, learning contribution is strongly concentrated toward the upper end of the initial-achievement distribution. The local contribution is substantially below average for small $x$, reaches the class-average level at $x \approx 0.58$, and exceeds the class average thereafter. The corresponding LCC is more strongly convex than the first model and lies further below the equal-contribution line, indicating a more pronounced concentration of learning gain among initially higher-achieving students.

**c) Linearly decreasing contribution.**

The third model reverses the pattern of the first model,

$$s(x) = 1.75 - 1.5x \tag{30}$$

which integrates to

$$y(x) = 1.75x - 0.75x^2 \tag{31}$$

In this case, $s(x) > 1$ for $x < 0.5$ and $s(x) < 1$ for $x > 0.5$. Students entering with lower initial achievement therefore exhibit above-average contribution, whereas students entering with higher initial achievement exhibit below-average contribution. The LCC is concave and lies above $y = x$, indicating that lower-ranked students account for more than their proportional share of the accumulated gain.

**d) Strongly decreasing contribution.**

The fourth model is

$$s(x) = 1 + 2.38(1 - 2x) \tag{32}$$

With corresponding LCC

$$y(x) = x + 2.38x(1 - x) \tag{33}$$

The LCP decreases strongly across the population. It is initially well above unity, crosses $s = 1$ at $x = 0.5$, and eventually becomes negative. Accordingly, the LCC initially rises rapidly, crosses $y = 1$, reaches a maximum of approximately $y_{max} \approx 1.20$, and subsequently decreases back to $y(1) = 1$. The portion of the population following the maximum therefore contributes negatively to the final net gain.

**e) Gaussian contribution.**

The final model represents a population in which learning contribution is concentrated near the middle of the initial-achievement distribution. This model is intentionally idealized, as an empirical LCP would not generally be expected to follow a Gaussian form. We specify

$$s(x) = 2.02\, exp\left[-\frac{(x - 0.5)^2}{2(0.2)^2}\right] \tag{34}$$

The corresponding LCC is obtained numerically through Eq. (18). The LCP is symmetric about $x = 0.5$ and reaches its maximum at the center of the initial-achievement distribution. Contributions are below the population average near both ends and above average across the central region. The resulting LCC is monotonic and S-shaped: it initially grows slowly, increases rapidly through the middle of the distribution, and then gradually approaches $y = 1$.

## 4. LC curves and Pre-post Score Structure

### 4.1 Regression model and the $\beta$ criterion

The illustrative models in Sec. 3 demonstrate a range of possible LCC and LCP shapes, but they do not reveal which statistical features of the pretest and posttest score distributions give rise to those patterns. In this section, we examine how the joint structure of the pretest and posttest distributions influences the learning contribution profiles, with particular emphasis on identifying the parameter that determines whether the LCP increases or decreases across initial achievement when individual normalized gain $g_i$ is used as the contribution variable $G_i$. Some notations used in this section should be introduced first. Let $P$ denote the pretest-score random variable and $p$ a particular value of $P$. We use $E[\cdot]$ to denote expectation, so that $E[g|P = p]$ represents the population mean normalized gain conditional on a pretest score $p$.

Because the LCP is defined in terms of cumulative pretest rank, let $R \in [0, 1]$ denote fractional rank according to pretest performance. The local LCP at a particular rank $R = x$ can then be written as

$$s(x) = \frac{E[g|R = x]}{E[g]} \tag{35}$$

Let $p(x)$ denote the pretest score associated with fractional rank $x$. Based on this definition, $p(x)$ is monotonically increasing function of $x$. Thus $E[g|R = x]$ is the essentially $E[g|P = p(x)]$. Then we have

$$s(x) = \frac{E[g|P = p(x)]}{E[g]} \tag{36}$$

Our approach to determine the shape of $s(x)$ hence relies on examining how the joint statistical relationship between pretest and posttest scores determines the conditional mean normalized gain $E[g|P = p]$. The central assumption adopted in this work is that the conditional mean of the posttest score $Q$ is approximately linear in the pretest score $P$:

$$E[Q|P = p] = \mu_Q + r\frac{\sigma_Q}{\sigma_P}(p - \mu_P) \tag{37}$$

Where $\mu_P$ and $\mu_Q$ are the population means, $\sigma_P$ and $\sigma_Q$ are the population standard deviations, and $r$ is the Pearson correlation between the pretest and posttest scores. Note that Eq. (37) applies provided that the pretest and posttest scores have finite second moments and that the pretest-score variance is strictly positive.

As a standard textbook result, if $P$ and $Q$ follow a bivariate normal distribution, the conditional mean of $Q$ given $P$ is linear [18], hence Eq. (37) applies[1]. Thus, the expected posttest score varies linearly with pretest score, with regression slope

$$\beta = r\frac{\sigma_Q}{\sigma_P} \tag{38}$$

Thus, the conditional mean can then be written using $\beta$ as

$$E[Q|P = p] = \mu_Q + \beta(p - \mu_P) \tag{39}$$

Note the coefficient $\beta$ combines two features of the pre-post relationship: the strength of the association between the two scores, represented by Pearson's $r$, and the relative dispersion of the posttest and pretest distributions, represented by $\sigma_Q/\sigma_P$.

Because the tie-invariant LCP is defined at the level of students sharing the same pretest score, the conditional mean above maps directly to the local-group mean normalized gain. For fixed $P = p$, this transformation is exact because $M - p$ is constant within the group

$$E[g|P = p] = \frac{E[Q|P = p] - p}{M - p} \tag{40}$$

Substituting the conditional mean of $Q$ gives

$$E[g|P = p] = \frac{\mu_Q - \beta\mu_P + (\beta - 1)p}{M - p} \tag{41}$$

This expression provides a direct connection between the joint pre-post score distribution and the local group level learning-gain profile. Within the continuous population model introduced above, $P$ and $Q$ are treated as continuous variables, and the conditional mean gain is differentiable for $p < M$. Differentiating with respect to pretest score therefore gives[2]

$$\frac{d}{dp}E[g|P = p] = \frac{\mu_Q - M + \beta(M - \mu_P)}{(M - p)^2} \tag{42}$$

Since the denominator $(M - p)^2$ is strictly positive, the direction in which the conditional normalized gain changes with initial achievement is determined entirely by the numerator,

$$\mu_Q - M + \beta(M - \mu_P) \tag{43}$$

Setting this quantity equal to zero defines a critical value

$$\beta_c = \frac{M - \mu_Q}{M - \mu_P} \tag{44}$$

The three resulting regimes are therefore

$$\begin{cases} \beta < \beta_c \Rightarrow \dfrac{dE[g|P = p]}{dp} < 0 \\ \beta = \beta_c \Rightarrow \dfrac{dE[g|P = p]}{dp} = 0 \\ \beta > \beta_c \Rightarrow \dfrac{dE[g|P = p]}{dp} > 0 \end{cases} \tag{45}$$

These results translate directly to the LCP. Recall $p(x)$ is a monotonically increasing function of $x$. For populations with positive overall mean individual gain, $\bar{g} > 0$. The differentiation of $s(x)$ gives

$$\frac{d}{dx}s(x) = \frac{1}{\bar{g}}\frac{dE[g|P = p(x)]}{dp}\frac{dp}{dx} \tag{46}$$

Since both $\bar{g}$ and $dp/dx$ are positive, the sign of $ds/dx$ is identical to the sign of $dE[g \mid P = p]/dp$. Therefore, $\beta < \beta_c$ indicates $s(x)$ decreases with initial achievement rank $x$, $\beta > \beta_c$ indicates $s(x)$ increases with initial achievement rank $x$. For the critical case where $\beta = \beta_c$, it is easily shown from Eq. (41) that

$$E[g|P = p] = \frac{\mu_Q - \mu_P}{M - \mu_P} = \bar{g} \tag{48}$$

This result indicates that the expected $g$ is independent of $p$. Thus, at the critical value, every region of the initial-achievement distribution has the same expected normalized gain, equal to the traditional Hake normalized gain calculated from the population mean pretest and posttest scores. The corresponding learning contributions are uniformly distributed across initial-achievement rank. Thus, $s(x) = 1$. Given the significance of $\beta$, we refer to it as the *score-structure parameter* and to $\beta_c$ as the *critical score-structure parameter* in the rest of the paper.

The corresponding LCC can be derived using Eq. (18). When $\beta < \beta_c$, learning contribution is relatively concentrated among students with lower initial achievement, producing a decreasing LCP and, under the monotonic conditions considered here, a concave LCC lying above the equal-contribution line $y = x$. When $\beta > \beta_c$, the pattern is reversed: learning contribution

---

[1] In practice, assessment scores are discrete and bounded, and their joint distribution may depart substantially from bivariate normality. Bivariate normality, however, is sufficient but not necessary for conditional linearity. The relevant assumption is the linearity of the conditional mean, whose empirical adequacy can be evaluated using a regression lack-of-fit test based on replicated observations at the same pretest scores.

[2] For a discrete assessment, the corresponding result is interpreted through differences between adjacent pretest-score groups rather than a literal derivative. Note for $M > p_2 > p_1$,

$$E[g|P = p_2] - E[g|P = p_1] = \frac{(p_2 - p_1)\left[\mu_Q - M + \beta(M - \mu_P)\right]}{(M - p_2)(M - p_1)}$$

Since all the other factors are positive, the same sign criterion holds exactly between discrete score groups.

becomes increasingly concentrated among students with higher initial achievement, the LCP increases, and the LCC becomes convex and lies below $y = x$. At the critical condition, the LCP is uniform, $s(x) = 1$, and the LCC coincides with the equal-contribution line, $y(x) = x$.

### 4.2 Simulated LCC and LCP across the $\beta$ regimes

To illustrate the three regimes predicted by the preceding analysis and Eq. (45), we generated five simulated pretest–posttest populations while holding the marginal score parameters fixed and varying the pre–post correlation. Each population contained $N = 10{,}000$ students on a 30-item assessment with $\mu_P = 10$, $\mu_Q = 17.5$, $\sigma_P = 2$, and $\sigma_Q = 3$. Thus, the critical value $\beta_c = 0.625$. We selected

$$\beta = 0.20, 0.40, 0.625, 0.80, 1.20$$

representing two cases with $\beta < \beta_c$, the critical case $\beta = \beta_c$, and two cases with $\beta > \beta_c$. Because $\beta = 3r/2$, these correspond to Pearson's $r$ of 0.13, 0.27, 0.42, 0.53, 0.80. respectively.

For each simulation, scores were generated from a latent bivariate-Gaussian model with the specified means, standard deviations, and correlation, then rounded to the nearest integer and constrained to the allowable score range. Posttest scores were bounded between 0 and 30, whereas pretest scores were bounded between 0 and 29 to ensure that the denominator $M - p$ in the normalized gain was nonzero. Each integer pretest score observed in the simulated sample defined one tied-score group, while score categories containing no students were omitted. Consequently, each of the five simulated populations contained between 15 and 17 tied-score groups. Individual normalized gains were then calculated, and the tie-invariant procedure described above was used to construct the LCC and LCP. Figure 2 shows the resulting LCCs and LCPs. As predicted, the LCP decreases with initial-achievement rank when $\beta < \beta_c$, approaches the equal-contribution line at $\beta = \beta_c$, and increases with rank when $\beta > \beta_c$. The corresponding LCCs lie predominantly above, near, and below the equal-contribution line $y(x) = x$. Thus, the simulations are consistent with the three regimes predicted by the analytical criterion Eq. (45).

The simulated LCPs exhibit small departures from the continuous prediction (see Sec. 4.3), particularly near the critical case. These departures arise from finite sampling and from rounding and bounding the simulated scores. Discretization makes the LCP stepwise, with sampling fluctuations most visible in sparsely populated pretest-score groups near the tails of the distribution. The LCC is less sensitive to these local fluctuations because it accumulates the contributions of successive groups.

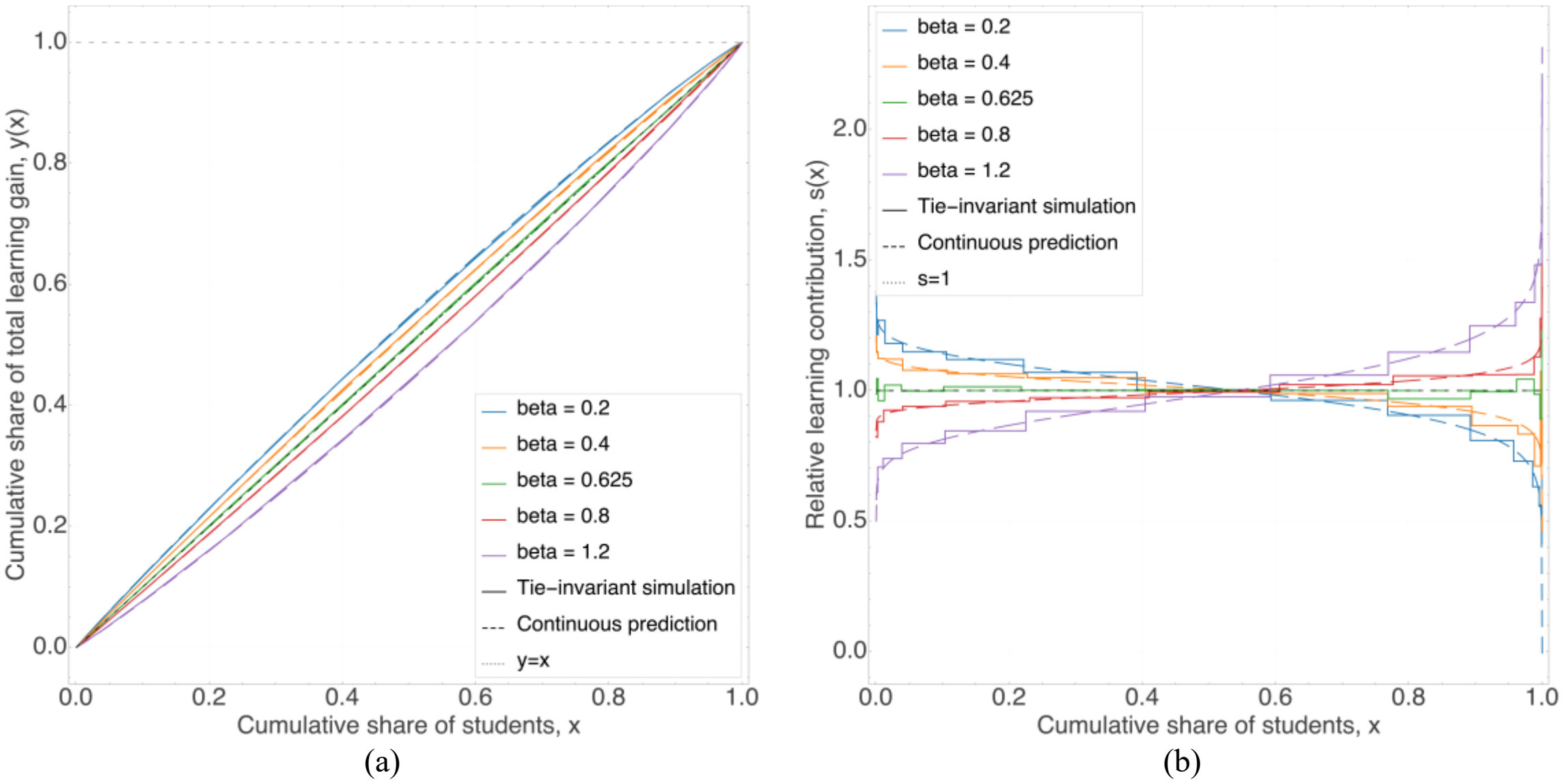

**FIG. 2.** Tie-invariant simulations and continuous score-structure predictions across five values of $\beta$, with $\beta_c = 0.625$.

### 4.3 Analytical prediction of LCP and LCC

Under the conditional linearity assumption, the parameter $\beta$ determines whether the expected normalized gain profile increases or decreases with initial achievement, but it does not by itself determine the complete shape of the LCP or LCC. To predict the curves analytically, the distribution of pretest scores must also be specified. This distinction is important because $\beta$ determines the direction of the gain pattern, whereas the pretest distribution determines how that pattern is represented across the cumulative rank.

Again, the model introduced in this section is predicting a local tie-group mean, not individual student learning. Recall equation (41) determines the expected normalized gain among a group of students with pretest score $p$. Using the expression for $\beta_c$, equation (41) can be rewritten as

$$E[g|P = p] = 1 - \beta_c + \frac{(\beta - \beta_c)(p - \mu_P)}{M - p} \tag{49}$$

When $\beta = \beta_c$, we will get $s(x) = 1$, and $y(x) = x$.

In the case where $\beta \neq \beta_c$, let's approximate the pretest distribution as Gaussian,

$$P \sim N(\mu_P, \sigma_P^2)$$

Let's again use $p(x)$ as the pretest score corresponding to the ranking score $x$. Under the Gaussian approximation, $x$ is the cumulative probability associated with $p(x)$:

$$x = \Phi\left(\frac{p - \mu_P}{\sigma_P}\right) \tag{50}$$

where $\Phi$ is the standard-normal cumulative distribution function. Equivalently,

$$p(x) = \mu_P + \sigma_P \Phi^{-1}(x) \tag{51}$$

Hence, we could rewrite Eq. (49) using Eq. (51),

$$E[g|P = p(x)] = 1 - \beta_c + \frac{(\beta - \beta_c)\sigma_P \Phi^{-1}(x)}{M - \mu_P - \sigma_P \Phi^{-1}(x)} \tag{52}$$

The predicted LCP is

$$s(x) = \frac{1 - \beta_c + \frac{(\beta - \beta_c)\sigma_P \Phi^{-1}(x)}{M - \mu_P - \sigma_P \Phi^{-1}(x)}}{\bar{g}} \tag{53}$$

Where the denominator is the mean individual gain:

$$\bar{g} = \int_0^1 \left(1 - \beta_c + \frac{(\beta - \beta_c)\sigma_P \Phi^{-1}(x)}{M - \mu_P - \sigma_P \Phi^{-1}(x)}\right) dx \tag{54}$$

The corresponding LCC can be obtained by Eq. (18). Using the parameters used for the simulation in Sec. 4.2, we were able to predict the LCC and LCP. The lines generated can be found in Fig. 2. as well. The analytical predictions generated from $s(x)$ and $y(x)$ closely reproduce the five simulated LCPs and LCCs across the full range of β values considered. The agreement is especially strong for the LCCs, while the small deviations observed in the LCPs are attributable primarily to finite sampling, score discretization, and the bounded nature of the simulated assessment. This close correspondence confirms that the analytical expressions for $s(x)$ and $y(x)$ reliably describe the distributional behavior observed in the simulations and provides a theoretical basis for interpreting the simulated LCC and LCP shapes.

The Gaussian formulation provides a convenient analytical approximation because it gives an explicit relationship between pretest score and the cumulative rank. An actual assessment, however, is bounded and discrete, whereas an unbounded Gaussian distribution formally permits scores outside the allowable range and assigns nonzero probability to values near $P = M$, where normalized gain becomes undefined. The continuous Gaussian model should therefore be interpreted as an idealized approximation when the score distribution lies sufficiently far from the assessment boundaries. A bounded or truncated distribution provides the corresponding exact formulation when boundary effects are important.

## 5 Application in a Classroom Study

### 5.1 Study Context

We applied the LC framework to matched responses from $N = 135$ preservice elementary teachers enrolled in a conceptual physics course at a university in the U.S. Midwest. The instructional unit focused on electric circuits. Student understanding was assessed using the DIRECT [5]. The version used in this study contains 29 MC questions, giving a maximum score of 29. The instructional period lasted approximately seven weeks. During this time, students attended one 50-minute lecture and one 170-minute laboratory session per week, with no additional course-related extracurricular activities. DIRECT was administered at the beginning of the instructional period and approximately seven weeks later using an individual-to-group sequence. Students first completed the assessment individually and then immediately worked on the same assessment in groups of three or four both times. These four conditions are denoted Pre-IN, Pre-G, Post-IN, and Post-G. During the group sessions, students were encouraged to discuss with each other but to provid their own responses and were not required to reach consensus; thus, group members did not necessarily submit identical answers. No teaching-assistant intervention occurred during these sessions. This design permits three complementary comparisons.

1) The transition Pre-IN→Pre-G characterizes the immediate change associated with collaborative work before instruction.
2) The transition Post-IN→Post-G characterizes the corresponding collaborative change after instruction.
3) The transition Pre-IN→Post-IN characterizes individual learning over the 7-week period.

### 5.2 An overview of the analysis

For each of the three transitions, we considered four representations of the learning-contribution pattern. First, the tie-invariant empirical LCC and LCP were constructed directly from the observed data using the procedure developed in the preceding theoretical section. Second, to summarize the overall empirical trend, we introduced a specific constrained empirical fitting model for the LCC and derived the corresponding fitted LCP analytically; this fitting procedure is described in detail below (see Sec. 5.3). Third, we developed a discrete score-structure prediction that uses the observed pretest-score groups together with the estimated pretest–posttest score parameters to predict the local and cumulative contribution patterns (see Sec. 5.4). Finally, a smooth theoretical LCC and LCP were generated by applying the population-level prediction developed in Sec. 4.3 to the score-structure parameters reported in Table I and II. Together, these four representations allow the directly observed contribution pattern, its empirical trend, and both discrete and continuous model-based predictions to be compared within a common framework. All four curves for each of the three transitions are shown in Figs. 3-5.

**Table I.** Descriptive statistics for the four DIRECT conditions. Standard deviations are sample standard deviations.

| Condition | Mean score $\mu$ | $\sigma$ |
|---|---|---|
| Pre-IN | 7.03 | 2.27 |
| Pre-G | 8.15 | 2.48 |
| Post-IN | 9.62 | 2.99 |
| Post-G | 11.87 | 2.80 |

**Table II.** Transition statistics, empirical fit results, and tests of the conditional-linearity assumption. Brackets give 95% bootstrap percentile intervals for $a$.

| Transition | $r$ | $\beta$ | $\beta_c$ | $a$ [95% bootstrap CI] | Lack of fit |
|---|---|---|---|---|---|
| Pre-IN to Pre-G | 0.30 | 0.33 | 0.95 | -4.84 [-11.77, -2.45] | $F(10,123) = 0.81, p = .62$ |
| Post-IN to Post-G | 0.56 | 0.52 | 0.88 | -1.96 [-3.18, -1.06] | $F(13,120) = 0.80, p = .66$ |
| Pre-IN to Post-IN | 0.22 | 0.29 | 0.88 | -1.83 [-2.93, -0.98] | $F(10,123) = 0.94, p = .50$ |

### 5.3 Empirical fitting of the LCCs and LCPs

To characterize the overall distributional trend and compare the empirical curves with the idealized models introduced in the preceding section, each tie-invariant LCC was fitted with the constrained quadratic function

$$y(x) = x + \frac{a}{2}(x^2 - x) \tag{55}$$

which satisfies the defining conditions $y(0) = 0$ and $y(1) = 1$. Importantly, the LCP was not fitted independently. Instead, the corresponding smooth LCP was obtained analytically from the derivative of the fitted LCC:

$$s(x) = \frac{dy}{dx} = 1 + a\left(x - \frac{1}{2}\right) \tag{56}$$

Uncertainty in $a$ was estimated using student-level bootstrap resampling. For each bootstrap replicate, $N = 135$ students were sampled with replacement from the original dataset, tied-score groups were reconstructed, the tie-invariant LCC was recalculated, and the constrained quadratic function was refitted. The resulting distribution of a was used to determine its bootstrap standard error and 95% confidence interval. The results of fitting can be found in TABLE II.

### 5.4 Discrete score structure and continuous prediction of the LCC and LCP

To examine whether the observed learning-contribution patterns could be predicted from the joint pretest–posttest score structure, we conducted two prediction analyses. First, we generated continuous predictions of the LCC and LCP for each of the three transitions using the score-structure parameter $\beta$. Following the model developed in Sec. 4.3, the pretest-score distribution was modeled as Gaussian using its observed mean and standard deviation. Second, we generated discrete score-structure predictions using $\beta$ and the observed pretest-score groups. We next describe the discrete prediction method in detail.

For each transition, the critical score-structure parameter $\beta$ and the critical score-structure parameter $\beta_c$ were reported in Table II. For all three transitions, $\beta < \beta_c$, predicting that the conditional mean normalized gain decreases with increasing initial achievement. The empirical LCP fits are consistent with this prediction.

To construct the discrete score-structure model for predicting the LCP and LCC of each transition, let $n_j$ denote the number of students in tied-score group $j$, whose members share the pretest score $P = p_j$. For a total of $K$ tied-score groups in the actual pretest score distribution, $j = 1, 2, \ldots, K$. The conditional mean post-score for this group was predicted as

$$E[Q|P = p_j] = \mu_Q + \beta(p_j - \mu_P) \tag{57}$$

Because all students within a tied score group share the same pretest score, this prediction maps directly to the predicted mean normalized gain of that group,

$$E[g|P = p_j] = \frac{E[Q|P = p_j] - p_j}{M - p_j} \tag{58}$$

The predicted LCP for each score group was therefore

$$s_j = \frac{E[g|P = p_j]}{\left(\sum_{j=1}^{K} n_j E[g|P = p_j]\right)/N} \tag{59}$$

Where the denominator is the predicted mean normalized gain $\bar{g}$. The corresponding group by group construction of LCC is constructed as

$$y_j = \frac{\sum_{l=1}^{j} n_l E[g|P = p_l]}{N\bar{g}} \tag{60}$$

Because both discrete and continuous predictions rely on the conditional-linearity assumption, we tested this assumption using replicated observations at each discrete initial score. The regression residual sum of squares was separated into pure-error and lack-of-fit components. No significant lack of fit was detected for any transition (Table II.). Thus, the data provided no evidence against the linear conditional-mean assumption for any of the three transitions.

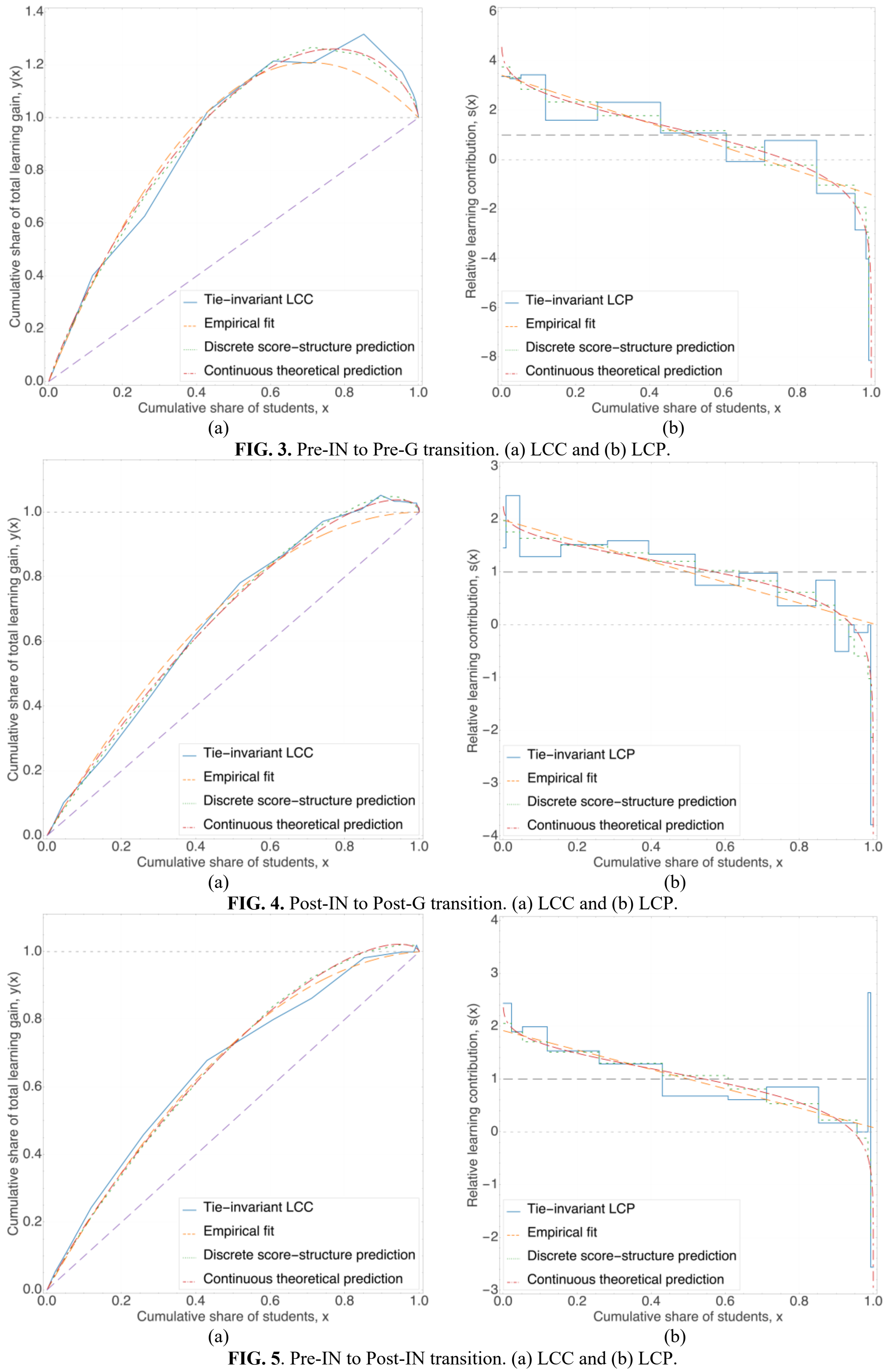

**FIG. 3.** Pre-IN to Pre-G transition. (a) LCC and (b) LCP.

**FIG. 4.** Post-IN to Post-G transition. (a) LCC and (b) LCP.

**FIG. 5**. Pre-IN to Post-IN transition. (a) LCC and (b) LCP.

### 5.5 Summary of the results

Across all three transitions, the empirical LCCs and LCPs reveal a consistent decrease in relative normalized gain with increasing initial achievement. This pattern is strongest for Pre-IN→Pre-G, followed by Post-IN→Post-G, and weakest for the seven-week Pre-IN→Post-IN transition. In the two individual-to-group transitions, the LCP becomes negative at the upper end of the achievement distribution, indicating negative average normalized gain for some higher-scoring groups. By contrast, the Pre-IN→Post-IN LCP remains positive across the full distribution.

The empirical fits provide a compact characterization of these overall trends. The fitted coefficient $a$ is negative for all three transitions, and its 95% bootstrap CI excludes zero in every case, providing consistent evidence of a systematic downward trend in relative normalized gain with increasing initial achievement. The fitted curves also distinguish the strength of this dependence across transitions, with the steepest decline before instruction in the Pre-IN→Pre-G comparison and the weakest decline across the seven-week instructional period.

The discrete score-structure predictions, generated from the observed pretest-score groups together with the estimated score structure parameter $\beta$, reproduce the empirical LCCs particularly well. The corresponding predicted LCPs also capture the general decreasing trend, although some deviations from the observed LCP are apparent at individual groups. Such deviations are expected because the empirical LCP reflects finite-sample variation within discrete tied-score groups, whereas the score-structure prediction represents their model-based conditional mean contribution.

The smooth continuous predictions provide a population-level representation of the same score-structure relationship. Across all three transitions, the predicted LCPs reproduce the observed direction and overall gradient of the contribution profiles, while the predicted LCCs closely capture the corresponding cumulative behavior. Taken together, the four representations are mutually consistent: the tie-invariant curves display the observed distributional structure, the empirical fits summarize its overall trend, and both the discrete and continuous score-structure predictions show that much of this structure can be anticipated from the means, standard deviations, and correlation of the paired score distributions. The remaining local departures, most visible in the empirical LCPs, represent structure not captured by the simple score-structure model.

Importantly, the uncertainty estimates for individual tied-score groups are not included in the empirical LCP. These local LCP values are sample means and may exhibit substantial sampling variability, particularly when a pretest score is represented by relatively few students. We therefore treat them primarily as descriptive features of the observed contribution profile rather than as independent inferential estimates. Statistical uncertainty in the overall distributional trend was instead assessed through student-level bootstrap resampling of the fitted LCC. Because tied-score groups were reconstructed within each bootstrap sample before refitting, this procedure propagates variation in local group composition into the uncertainty of the fitted parameter.

## 6 Discussion

The LC framework developed in this work extends conventional class-level summaries by characterizing how the observed learning contribution is distributed across the initial-achievement spectrum. The framework introduces an intermediate group-level perspective based on students with the same initial score. Starting from an individual contribution measure $G$, students with similar initial achievement are considered collectively through their local mean contribution, which is then compared with the overall mean individual gain. This local aggregation reduces emphasis on potentially noisy individual-level variation while retaining systematic differences associated with initial achievement. The framework formalizes this individual-to-group-to-class relationship through two complementary representations: the LCP expresses the local-group mean contribution relative to the class mean individual contribution, whereas the LCC accumulates these relative contributions across the initial-achievement distribution. When individual Hake normalized gain $g$ is used as $G$, the local group quantity has an especially direct interpretation: for students sharing the same pretest score, their mean individual normalized gain is mathematically identical to the normalized gain calculated from the local group mean scores. The framework therefore connects individual score changes, local group behavior, and class-level outcomes within a common distributional representation. Moreover, as demonstrated by the score-structure analysis, the local contribution pattern can be related quantitatively to the joint statistical structure of the pretest and posttest scores through the score-structure parameter $\beta$ and the critical score-structure parameter $\beta_c$. The relation between $\beta$ and $\beta_c$ determines whether the predicted local contributions increase, decrease, or remain constant across the initial-achievement spectrum. Thus, the analysis provides not only a descriptive account of where observed learning gains are concentrated but also a model-based reference for assessing how much of the observed LCC and LCP structure can be anticipated from the underlying joint score distribution.

The LCC and LCP supplement class-average measures by revealing distributional features that a single average cannot capture. The class-average normalized gain summarizes the overall magnitude of learning, whereas the LCC shows how the total observed learning gain accumulates across students ordered by initial achievement, thereby revealing whether gains are concentrated among students with lower, middle, or higher initial scores. The LCP provides a corresponding local perspective by identifying regions of the initial-achievement spectrum in which students make above-average, below-average, or negative contributions to the total gain. The framework can therefore reveal heterogeneity, including localized learning losses and compensating gains, that is obscured by the class average. Two classes with the same class-average gain may exhibit substantially different LCCs and LCPs; conversely, classes with different average gains may have similar contribution patterns. Reporting the class-average gain together with the LCC and LCP therefore characterizes both the overall magnitude and the distribution of learning.

Care is required when comparing LCCs or LCPs across populations for two reasons. First, the horizontal coordinate $x$ represents relative rank within each population rather than an absolute level of prior achievement. Thus, the same value of $x$ may correspond to different pretest scores in different populations. Curve shapes are therefore most directly comparable when the populations have similar initial-achievement distributions. When these distributions differ substantially, comparisons at a common $x$ should be interpreted as comparisons between students at similar relative positions, not necessarily students with equivalent prior preparation. Research questions concerning comparable absolute levels of initial achievement should instead focus on overlapping pretest-score ranges or use methods that adjust for baseline differences. Second, the LCP represents relative contribution because $s(x) = G(x)/\bar{G}$, where $G(x)$ is the local group mean, and $\bar{G}$ is the population mean of the individual learning measure. Equal values of $s(x)$ in two populations therefore indicate the same contribution relative to each population's own mean, not the same learning gain on the original scale of the measure. The local mean can be recovered from $s(x)\bar{G}$. For example, if the individual learning measure is the Hake normalized gain, $s(x)\bar{g}$ gives the corresponding local group mean normalized gain. Cross-population comparisons should therefore report the LCC and LCP together with Hake's class average gain to characterize both the distribution and the overall magnitude of learning.

Similar caution applies to causal interpretation. An LCP showing larger gains among lower-pretest students describes an observed distributional pattern but does not establish that instruction caused greater learning for this group. Such claims require an appropriate research design, such as random assignment, or statistical adjustment for baseline differences and other confounding factors. Thus, the LCC and LCP characterize observed learning contributions, whereas causal interpretation depends on the study design.

**Limitations of the study**

Several limitations should be recognized. First, the framework inherits the statistical and measurement properties of the chosen contribution measure, $G$. In the present study, using individual normalized gain leaves concerns about pretest dependence, floor and ceiling effects, measurement uncertainty, guessing, and test-taking behavior. Although the local averaging underlying the LCP can reduce idiosyncratic student-level fluctuations, it cannot eliminate systematic features of the measurement process. Second, because assessment scores are discrete, ties in initial scores are common. The tie-invariant procedure eliminates dependence on arbitrary within-tie ordering by assigning each student in a tied-score group the group's mean contribution. This produces well-defined LCC and LCP segments but does not display variation within tied-score groups. Third, LCC normalization requires a nonzero total contribution, and the present formulation is restricted to populations with a positive mean contribution. When the mean is close to zero, normalization can magnify modest local differences and produce unstable curves. Populations with zero or negative mean contributions therefore require a modified interpretation or an alternative formulation.

**Future directions**

Several directions for future research follow. Applying the learning-contribution framework across conceptual assessments, instructional settings, and student populations could determine whether recurring distributional patterns generalize across courses and disciplines. Appropriately designed comparative studies could examine differences in LCC and LCP structure across instructional conditions or preparation-based subgroups while accounting for baseline differences. Methodological work could develop confidence bands and global tests for comparing entire curves, as well as flexible estimation methods for empirical profiles not adequately represented by the illustrative models considered here. Robustness studies could also examine alternative contribution measures, including raw gain, model-based estimates of change, and measures designed to address limitations of normalized gain. Finally, connecting the LCC to the broader literature on concentration curves may provide additional methods for statistical inference and summary while preserving its learning-specific interpretation.

## Acknowledgements

This work is supported in part by the U.S. National Science Foundation grant 2111138. Opinions expressed are of the author and not necessarily of the Foundation.